\documentclass[a4paper,11pt]{article}

\usepackage{jheppub}
\usepackage{graphicx}
\usepackage{amsmath}
\usepackage{slashed}
\usepackage{braket}
\usepackage[export]{adjustbox}
\usepackage{enumitem}
\usepackage{placeins}
\usepackage[normalem]{ulem}

\newcommand{\refcite}[1]{ref.~\cite{#1}}
\newcommand{\refscite}[1]{refs.~\cite{#1}}
\newcommand{\Eq}[1]{Eq.~\eqref{eq:#1}}
\newcommand{\eq}[1]{eq.~\eqref{eq:#1}}
\newcommand{\eqs}[2]{eqs.~\eqref{eq:#1} and \eqref{eq:#2}}
\renewcommand{\sec}[1]{section~\ref{sec:#1}}

\newcommand{\app}[1]{appendix~\ref{app:#1}}
\newcommand{\fig}[1]{figure~\ref{fig:#1}}

\newcommand{\df}{\mathrm{d}}
\newcommand{\img}{\mathrm{i}}
\newcommand{\eps}{\epsilon}

\newcommand{\bn}{{\bar n}}

\newcommand{\GeV}{\,\mathrm{GeV}}
\newcommand{\cI}{\mathcal{I}}

\newcommand{\cL}{\mathcal{L}}
\newcommand{\cM}{\mathcal{M}}
\newcommand{\cN}{\mathcal{N}}
\newcommand{\cO}{\mathcal{O}}
\newcommand{\Tau}{\mathcal{T}}

\newcommand{\wa}{{w_1}}
\newcommand{\wb}{{w_2}}
\newcommand{\cut}{{\rm cut}}
\newcommand{\nlim}{\lim\limits_{\mathrm{strict}~n-\mathrm{coll.}}}
\newcommand{\as}{\alpha_s}
\newcommand{\GammaC}{\Gamma_{\rm cusp}}
\newcommand{\nn}{\nonumber}
\newcommand{\MSbar}{\overline{\mathrm{MS}}}
\newcommand{\lqcd}{\Lambda_\mathrm{QCD}}

\def\beq{\begin{equation}}
\def\eeq{\end{equation}}
\def\bea{\begin{eqnarray}}
\def\eea{\end{eqnarray}}

\title{\boldmath $N$-jettiness beam functions at N$^3$LO}

\author[a]{Markus A.~Ebert,}
\emailAdd{ebert@mit.edu}

\author[b]{Bernhard Mistlberger,}
\emailAdd{bernhard.mistlberger@gmail.com}

\author[a]{and Gherardo Vita}
\emailAdd{vita@mit.edu}

\affiliation[a]{Center for Theoretical Physics, Massachusetts Institute of Technology, Cambridge, Massachusetts 02139, USA}
\affiliation[b]{SLAC National Accelerator Laboratory, Stanford University, Stanford, CA 94039, USA}

\abstract{
We present the first complete calculation for the quark and gluon $N$-jettiness ($\Tau_N$) beam functions at next-to-next-to-next-to-leading order (N$^3$LO) in perturbative QCD.
Our calculation is based on an expansion of the differential Higgs boson and Drell-Yan production cross sections about their collinear limit.
This method allows us to employ cutting edge techniques for the computation of cross sections to extract the universal building blocks in question.
The class of functions appearing in the matching coefficents for all channels includes iterated integrals with non-rational kernels, thus going beyond the one of harmonic polylogarithms.
Our results are a key step in extending the $\Tau_N$ subtraction methods to N$^3$LO, and to resum $\Tau_N$ distributions at N$^3$LL$^\prime$ accuracy both for quark as well as for gluon initiated processes.
}

\preprint{MIT-CTP 5208}

\begin{document}

\maketitle

\section{Introduction}
\label{sec:intro}

Experimental measurements at the LHC have provided remarkably precise measurements
for a multitude of observables, most notably weak gauge boson production, an important
benchmark for the Standard Model which has been measured at percent level accuracy
\cite{Aaboud:2017svj,Aaboud:2017ffb,Khachatryan:2016nbe,Sirunyan:2019bzr}.
Strong constraints on physics beyond the Standard Model are also provided by
precision measurements of Higgs boson production and diboson processes
\cite{Sirunyan:2018koj,Aad:2020mkp,Aaboud:2019nkz,Aaboud:2019lgy,Sirunyan:2019gkh}.
To make full use of these results, it is crucial to confront them with equally-precise theory predictions,
which in particular requires to include higher-order corrections in QCD.

So far, only inclusive Drell-Yan and Higgs production have been calculated at next-to-next-to-next-to-leading order (N$^3$LO) in QCD~\cite{Anastasiou:2014vaa,Anastasiou:2015ema,Dreyer:2016oyx,Dreyer:2018qbw,Mistlberger:2018etf,Duhr:2019kwi,Duhr:2020kzd,Duhr:2020seh}, while significant progress is being made to reach the same precision for differential distributions~\cite{Cieri:2018oms,Dulat:2018bfe}.
A key challenge for such calculations is the cancellation of infrared divergences
between real and virtual corrections, and hence a necessary prerequisite is a profound
understanding of the infrared singular structure at three loops.

$N$-jettiness ($\Tau_N$) is an infrared-sensitive $N$-jet resolution observable
and thus provides a way to study the singular structure of QCD~\cite{Stewart:2009yx, Stewart:2010tn}.
Its simplest manifestation $\Tau_0$, also referred to as beam thrust, is defined as
\begin{align} \label{eq:Tau0_0}
 \Tau_0 = \sum_i \min \biggl\{ \frac{2 q_a \cdot k_i}{Q_a} \,, \frac{2 q_b \cdot k_i}{Q_b} \biggr\}
\,,\end{align}
where the sum rums over all momenta $k_i$ in the hadronic final state,
$q_{a,b}$ are the momenta of the incoming partons projected onto the Born kinematics,
and the measures $Q_{a,b}$
distinguish different definitions of $\Tau_0$~\cite{Stewart:2010pd, Jouttenus:2011wh}.
A key feature of $\Tau_N$ is that its singular structure as $\Tau_N\to0$ is fully captured by a factorization theorem, as shown in~\refscite{Stewart:2009yx, Stewart:2010tn} using soft-collinear effective theory (SCET)~\cite{Bauer:2000ew,Bauer:2000yr,Bauer:2001ct,Bauer:2001yt,Bauer:2002nz}.
In the simplest case, namely the production of a color-singlet final state $h$, the appropriate factorization theorem reads
\begin{align} \label{eq:Tau0_factorization}
 \frac{\df\sigma}{\df Q^2 \df Y \df \Tau_0} &
 = \sigma_0 \sum_{a,b} H_{ab}(Q^2, \mu) \int \! \df t_a \, \df t_b \,
   B_a(t_a, x_a, \mu) \, B_b(t_b, x_b, \mu) \,
   S_c\Bigl(\Tau_0 - \frac{t_a}{Q_a} - \frac{t_b}{Q_b}, \mu\Bigr)
\nn \\ &\quad
\times \Bigl[1 + \cO\Bigl(\frac{\Tau_0}{Q}\Bigr)\Bigr]
\,.\end{align}
Here, $Q^2$ and $Y$ are the invariant mass and rapidity of $h$, respectively,
and we normalize by the Born partonic cross section $\sigma_0$.
In \eq{Tau0_factorization}, the full process dependence is given in terms of the hard function $H_{ab}$,
which encodes virtual corrections to the underlying hard process $a b \to h$.
The beam functions $B_{a,b}$ encode radiation collinear to the incoming partons.
The soft function $S_c$ encodes soft radiation and only depends on the color channel $c\in\{gg,q\bar q\}$,
but is independent of quark flavors.
Both beam and soft functions are universal and process independent.
Since they are defined as gauge-invariant matrix elements in SCET, calculating them at higher orders also provides a well-defined means of separately studying the collinear and soft limits of QCD themselves.
The beam functions $B_{a,b}$ not only appear in the factorization theorem for all $\Tau_N$, but also arise in the factorization theorem  for the generalized threshold inclusive color-singlet production in hadronic collisions~\cite{Lustermans:2019cau}, and are thus of particular interest on their own.

Since \eq{Tau0_factorization} fully captures the singular limit of QCD,
it can be employed as a subtraction scheme for higher-order calculations~\cite{Boughezal:2015dva,Gaunt:2015pea},
in analogy to the $q_T$ subtraction method based on a similar factorization for the transverse-momentum distribution \cite{Catani:2007vq}.
For both methods, extensions to N$^3$LO have been recently proposed~\cite{Cieri:2018oms,Billis:2019vxg}.
The $\cO(\Tau_0/Q)$ corrections to \eq{Tau0_factorization} have also been studied in the context of $\Tau_N$ subtractions \cite{Moult:2016fqy, Moult:2017jsg, Ebert:2018lzn, Boughezal:2016zws, Boughezal:2018mvf, Boughezal:2019ggi}.%
\footnote{For measurements with fiducial cuts applied to $h$, \eq{Tau0_factorization} also receives enhanced $\cO(\sqrt{\Tau_0/Q})$ corrections~\cite{Ebert:2019zkb}.}
These calculations are also interesting on their own as they provide insights into the infrared structure of QCD beyond leading power.
$\Tau_0$ subtractions are also the basis of combining NNLO calculations with a parton shower in \texttt{GENEVA}~\cite{Alioli:2012fc,Alioli:2015toa}.

Currently, the quark and gluon $\Tau_N$ beam functions are known at NNLO~\cite{Stewart:2010qs,Berger:2010xi,Gaunt:2014xga,Gaunt:2014cfa}, and significant progress has been made towards the calculation at N$^3$LO for the quark case~\cite{Melnikov:2018jxb, Melnikov:2019pdm,Behring:2019quf}.
The soft functions required for $\Tau_{0,1,2}$ are known at NNLO~\cite{Schwartz:2007ib, Fleming:2007xt, Kelley:2011ng, Monni:2011gb, Hornig:2011iu, Boughezal:2015eha,Campbell:2017hsw, Jin:2019dho}.
The factorization for $\Tau_{N\ge1}$ also requires the so-called jet function,
which is also known at N$^3$LO~\cite{Bauer:2003pi,Becher:2006qw,Fleming:2003gt,Becher:2009th,Becher:2010pd,Bruser:2018rad,Banerjee:2018ozf}.
In this paper, we calculate the $\Tau_N$ beam functions for all partonic channels at N$^3$LO,
thereby providing a critical ingredient to extending $\Tau_N$ subtraction to three loops both for quark as well as for gluon initiated processes.

Our computation is based on a method of expanding cross sections around the kinematic limit in which all final state radiation becomes collinear to one of the scattering hadrons~\cite{Ebert:CollExp}.
This method allows one to efficiently connect technology for the computation of scattering cross sections to universal building blocks of perturbative QFT.
In particular, we perform a collinear expansion of the Drell-Yan and gluon fusion Higgs boson production cross section at N$^3$LO.
Subsequently, we employ the framework of reverse unitarity~\cite{Anastasiou2003,Anastasiou:2002qz,Anastasiou:2003yy,Anastasiou2005,Anastasiou2004a}, integration-by-part (IBP) identities~\cite{Chetyrkin:1981qh,Tkachov:1981wb} and the method of differential equations~\cite{Kotikov:1990kg,Kotikov:1991hm,Kotikov:1991pm,Henn:2013pwa,Gehrmann:1999as} to obtain the collinear limit of the cross sections differential in the rapidity and transverse momentum of the colorless final states.
Using the connection of this limit to the desired beam functions we extract the desired perturbative matching kernels as discussed in \refcite{Ebert:CollExp}.

This paper is structured as follows.
In \sec{coll_exp}, we discuss our setup for calculating the beam functions based on the collinear expansion of \refcite{Ebert:CollExp}.
In \sec{results}, we briefly present our results, before concluding in \sec{conclusion}.
Our results are also provided in the form of ancillary files with this submission.

\section{Beam functions from the collinear limit of cross sections}
\label{sec:coll_exp}

Since the $\Tau_N$ beam function is independent of $N$, we calculate it from the simplest case $\Tau_0$ by considering the production of a colorless hard probe $h$
and an additional hadronic state $X$ in a proton-proton collision,
\begin{align} \label{eq:process_hadr}
 P(P_1) + P(P_2) \quad\to\quad h(-p_h) + X(-k)
\,,\end{align}
where the incoming protons are aligned along the directions
\begin{align} \label{eq:nnb}
 n^\mu = (1,0,0,1) \,,\qquad \bn^\mu = (1,0,0,-1)
\end{align}
and carry the momenta $P_1$ and $P_2$ with the center of mass energy $S=(P_1+P_2)^2$. The hard probe $h$ carries the momentum $p_h$, and the total momentum of the hadronic final state is denoted as $k$. We parameterize these momenta in terms of
\begin{align} \label{eq:vardef}
 Q^2 = p_h^2
 \,, \quad
 Y = \frac12 \ln\frac{\bn \cdot p_h}{n \cdot p_h}
 \,, \quad
 \wa = -\frac{\bar n\cdot k}{\bar n \cdot p_1}
 \,,\quad
 \wb = -\frac{ n\cdot k}{ n \cdot p_2}
 \,,\quad
 x = \frac{k^2}{(\bar n\cdot k)(n\cdot k)}
\,,\end{align}
where $Q^2$ and $Y$ are the invariant mass and rapidity of the hard probe $h$, respectively.

\Eq{process_hadr} receives contributions from the partonic process
\begin{align} \label{eq:process_part}
 i(p_1) + j(p_2) \quad\to\quad h(-p_h) + X_n(-p_3, \dots, -p_{n+2})
\,,\end{align}
where $i$ and $j$ are the flavors of the incoming partons which carry the momenta $p_1$ and $p_2$, and $X_n$ is a hadronic final state consisting of $n$ partons with the momenta $\{-p_3, \dots, -p_{n+2}\}$, and $n=0$ at tree level.
The cross section for the partonic process in \eq{process_part}, differential in the variables defined in \eq{vardef}, is then defined as
\begin{align} \label{eq:sigma_part}
  \frac{\df \eta_{ij}}{ \df Q^2  \df \wa \df \wb \df  x}  &
 = \frac{1}{\sigma_0} \frac{\cN_{ij}}{2 S} \sum_{X_n}\int  \frac{\df\Phi_{h+n} }{ \df \wa \df \wb \df  x}\, |\cM_{ij\to h+X_n}|^2
\,.\end{align}
Here, we normalize by the partonic Born cross section $\sigma_0$,
$\df\Phi_{h+n}$ is the phase space measure of the $h+X_n$ state,
and $|\cM_{ij\to h+X_n}|^2$ is the squared matrix element for the process in \eq{process_part},
summed over the colors and helicities of all particles, with $\cN_{ij}$ accounting for the color and helicity average of the incoming particles. Explicit expressions for $\cN_{ij}$ and $\df\Phi_{h+n}$ can be found in \refcite{Ebert:CollExp}.

The partonic cross section in \eq{sigma_part} is closely related to the beam function we are interested in.
For perturbative values of $\Tau_N$, one can match the beam functions onto the PDFs as \cite{Stewart:2009yx, Stewart:2010qs}
\begin{equation} \label{eq:beam_OPE}
 B_i(t,z,\mu)
 = \sum_j \int_z^1 \frac{\df z'}{z'} \, \cI_{ij}(t,z',\mu) \, f_j\Bigl(\frac{z'}{z},\mu\Bigr)
 \times \biggl[1+\mathcal{O}\biggl(\frac{\lqcd^2}{t}\biggr)\biggr]
\,.\end{equation}
Here, $\cI_{ij}$ is a perturbative matching kernel, and $t = \Tau_0 Q_a$, see \eq{Tau0_factorization}.
As shown by us in \refcite{Ebert:CollExp}, $\cI_{ij}$ is precisely given by the strict $n$-collinear limit of \eq{sigma_part}, where all loop and real momenta are treated as being collinear to $n$-direction, and we refer to \refcite{Ebert:CollExp} for details on how to calculate this limit:
\begin{align} \label{eq:Iij_bare}
 \cI_{ij}(t,z,\eps)
 = \int_0^1 \df x \int_0^\infty \df \wa\df \wb \,&
   \delta[z - (1-\wa)] \delta(t - Q^2 \wb)
   \nn\\&\times\,
   \nlim \frac{\df\eta_{ij}}{\df Q^2  \df\wa \df\wb \df x}
\,.\end{align}
Here, we have regulated both UV and IR divergences by working in $d=4-2\eps$ dimensions. The renormalized matching kernel is then given by~\cite{Stewart:2010qs,Berger:2010xi,Ebert:CollExp}
\begin{equation} \label{eq:Iij_ren}
 \cI_{ij}(t,z,\mu)
 = \sum_{k} \int\df t' \, Z_B^i(t-t',\eps,\mu) \, \int_z^1\!\frac{\df z'}{z'}
   \, \Gamma_{jk}\Bigl(\frac{z}{z'},\eps\Bigr) \,
   \hat Z_{\as}(\mu,\eps) \, \cI_{i k}(t',z',\eps)
\,.\end{equation}
Here, $\hat Z_{\as}$ implements the standard UV renormalization by renormalizing the bare coupling constant $\as^b$ in the $\MSbar$ scheme, and the convolution with the PDF counterterm $\Gamma_{jk}$ cancels infrared divergences. Explicit expressions for these ingredients are collected in \app{UV_IR_counter_terms}.
The remaining poles in $\eps$ are canceled by the convolution with the beam function counter term $Z_B$,
which in the formulation of the beam function within SCET arises as an additional UV counter term in effective theory.

\section{Results}
\label{sec:results}

In this section we report on our results for the matching kernels through N$^3$LO.
Our computation is based on the collinear expansion of the cross sections for the production of a Higgs boson via gluon fusion and for the production of off-shell photon (Drell-Yan) in hadron collisions. 
We compute the Higgs boson production cross section in the heavy top quark effective theory where the degrees of freedom of the top quark were integrated out and the Higgs boson couples directly to gluons~\cite{Inami1983,Shifman1978,Spiridonov:1988md,Wilczek1977,Chetyrkin:1997un,Schroder:2005hy,Chetyrkin:2005ia,Kramer:1996iq}.

We begin by computing all required matrix elements with at least one final state parton to obtain N$^3$LO cross sections. 
All partonic cross sections corresponding to matrix elements with exactly one parton in the final state were obtained in full kinematics for the purpose of \refscite{Duhr:2020seh,Dulat:2014mda,Dulat:2017brz,Dulat:2017prg} and are in part based on \refscite{Anastasiou:2013mca,Duhr:2014nda,Duhr:2013msa}.
In order to obtain the strict collinear limit we simply expand the existing results and select the required components.

To compute partonic cross sections with more than one final state parton we generate the necessary Feynman diagrams using QGRAF~\cite{Nogueira_1993} and perform spinor and color algebra in a private code.
Subsequently, we perform the strict collinear expansion of this matrix elements as outlined in \refcite{Ebert:CollExp}.
We make use of the framework of reverse unitarity ~\cite{Anastasiou2003,Anastasiou:2002qz,Anastasiou:2003yy,Anastasiou2005,Anastasiou2004a} in order to integrate over loop and phase space momenta.
We apply integration-by-part (IBP) identities~\cite{Chetyrkin:1981qh,Tkachov:1981wb} in order to re-express our expanded cross section in terms of collinear master integrals depending on the variables introduced in \eq{vardef}.
We then compute the required master integrals using the method of differential equations~\cite{Kotikov:1990kg,Kotikov:1991hm,Kotikov:1991pm,Henn:2013pwa,Gehrmann:1999as}.
In order to fix all boundary conditions for the differential equations we expand the collinear master integrals further around the soft limit and integrate over phase space.
The result of this procedure is then easily matched to the soft integrals that were obtained for the purpose of \refscite{Anastasiou:2013srw,Anastasiou:2014lda,Anastasiou:2014vaa,Anastasiou:2015yha,Duhr:2019kwi}.

This yields all required ingredients for the bare partonic cross section expanded in the strict collinear limit of \eq{sigma_part}.
This part of the computation is the same as for the results of \refcite{Ebert:PTBF}.
Next, we perform the Fourier transform over $t$ and make use of \eq{Iij_bare} to obtain the  matching kernel through N$^3$LO in QCD perturbation theory.
We will elaborate on the details of the computation of the matching kernels in \refcite{Ebert:ThingsToCome}.
Finally, we subtract poles in $\eps$ as given in \eq{Iij_ren} to obtain the renormalized matching kernel $ \cI_{ij}(t,z,\mu)$ through N$^3$LO in QCD perturbation theory. This is carried out in Fourier ($y$) space, where the convolution in $t$ becomes a simple product, and the Fourier-transformed counter term $\tilde Z_B$ can be easily predicted from the known renormalization group equation (RGE) of the beam function. We collect the required formulas in \app{counter_term}. It straightforward to Fourier transform back to $t$ space after the UV renormalization, and we will provide results in both spaces.

We express the perturbative matching kernels in terms of harmonic polylogarithms~\cite{Remiddi:1999ew} and Goncharov polylogarithms~\cite{Goncharov:2001iea} as well as a set of iterated integrals.
We define the iterated integrals recursively via
\beq
J\bigl(a_1(z),a_2(z),\dots,a_n(z)\bigr)=\int_0^z \df x \, a_1(x) \, J\bigl(a_2(x),\dots,a_n(x)\bigr)\,,
\eeq
with the prescription to regularize logarithmic singularities as
\beq
J\left(\frac{1}{z}\right)=\int_1^z \frac{dx}{x}=\log(z).
\eeq
We refer to the arguments of the iterated integrals as letters. 
The explicit end point of the iterated integration used for our iterated integrals is always the variable $\bar z=1-z$.
In order to express our matching kernels we require the following set of letters (or alphabet):
\beq
\mathcal{A}=\left\{\frac{1}{z},\frac{1}{1-z},\frac{1}{2-z},\frac{1}{1+z},\frac{1}{z},\frac{1}{\sqrt{4-z} \sqrt{z}}\right\}\,.
\eeq
It is possible to rationalise the square root in $\mathcal{A}$ by introducing the variable transformation $z\to (y+1)^2/y$ as noted in \refcite{Behring:2019quf} and to rewrite the iterated integrals in terms of Goncharov polylogarithms using well known techniques, see for example \refscite{Duhr:2011zq,Duhr:2012fh,Duhr:2019tlz,Panzer:2014caa}.

Studying the letters of our alphabet and the singularities appearing in our matching kernels  we see that they contain logarithmic singularities at the boundaries of the physical interval $z\in [0,1]$. 
In order to provide a representation of our perturbative matching kernels that is suitable for numeric evaluation we perform a generalised power series expansion around two different points $z=0$ and $z=1$ up to 50 terms in the expansion. 
Both power series are formally convergent within the entire unit interval but converge of course faster if the respective expansion parameter is smaller.
We provide both power series for all matching kernels as well as the analytic solution in ancillary files together with the arXiv submission of this article.

We have calculated the matching kernel in Fourier ($y$) space, where its renormalization becomes simpler.
As it is more commonly used in momentum ($t$) space, we provide results in both spaces.
The corresponding kernels are expanded in powers of $\as/\pi$,
\begin{align} \label{eq:I_as_expansion}
 \tilde\cI_{ij}(y,z,\mu) &= \sum_{n=0}^\infty \Bigl(\frac{\as}{\pi}\Bigr)^n \tilde\cI_{ij}^{(n)}(y,z,\mu)
\,,\qquad
 \cI_{ij}(t,z,\mu) = \sum_{n=0}^\infty \Bigl(\frac{\as}{\pi}\Bigr)^n \cI_{ij}^{(n)}(t,z,\mu)
\,.\end{align}
The coefficients $\tilde\cI_{ij}^{(n)}$ and $\cI_{ij}^{(n)}$ can be further expanded as
\begin{align} \label{eq:I_log_expansion}
 \tilde\cI_{ij}^{(n)}(y,z,\mu) &= \tilde I_{ij}^{(n)}(z) + \sum_{m=1}^{2n} \tilde\cI_{ij}^{(n,m)}(z) L_y^m
\,,\nn\\
 \cI_{ij}^{(n)}(t,z,\mu) &= \delta(t) I_{ij}^{(n)}(z) + \sum_{m=0}^{2n-1} \cI_{ij}^{(n,m)}(z) \cL_m(t,\mu^2)
\,,\end{align}
where the logarithm $L_y$ and the distribution $\cL_m$ are defined as
\begin{align}
 L_y = \ln\bigl(\img y \mu^2 e^{\gamma_E}\bigr)
\,,\qquad
 \cL_m(t,\mu^2) = \biggl[ \frac{1}{t} \ln^m\frac{t}{\mu^2}\biggr]_+
\,,\end{align}
where the $[\cdots]_+$ denotes the standard plus distribution.
Note that there is no one-to-one correspondence between the $\tilde \cI_{ij}^{(\ell,m)}(z)$ and $ \cI_{ij}^{(\ell,m)}(z)$, as the Fourier transform induces a nontrivial mixing. 
For explicit relations for the Fourier transform, see e.g.~\refcite{Ebert:2016gcn}.

The logarithmic terms in \eq{I_log_expansion} encode the scale dependence of the beam function, and thus their structure is fully determined by its renormalization group equation (see \app{RGE}) in terms of its anomalous dimensions and lower-order ingredients.
The genuinely new three-loop results calculated by us are the nonlogarithmic boundary terms $\tilde I_{ij}^{(3)}(z)$ and $I_{ij}^{(3)}(z)$.

We performed several checks on our results. Firstly, we verified that all poles in $\eps$ cancel after applying UV renormalization and IR subtraction as given in \eq{Iij_ren}, where the beam function counterterm was predicted from its RGE as shown in \app{counter_term}.
To check that our results obey the beam function RGE, we verified all logarithmic terms in \eq{I_log_expansion} against those predicted in \refcite{Billis:2019vxg} by solving the beam function RGE.
We also checked that our results for $I_{ij}(z)$ agree with the eikonal limit $\lim_{z\to1} I_{ij}^{(3)}(z)$ that was predicted in \refcite{Billis:2019vxg} using a consistency relation with the threshold soft function \cite{Lustermans:2019cau}, and that our results agree with the generalized large-$n_c$ approximation $n_c \sim n_f \gg 1$ obtained in \refcite{Behring:2019quf}.
Furthermore, we checked that the first four terms in the soft expansion of the Higgs boson production cross section reproduce correctly the collinear limit of the threshold expansion of the partonic cross section obtained for the purpose of \refscite{Dulat:2017prg,Dulat:2018bfe}.
The inclusive cross section at N$^3$LO for Drell-Yan and Higgs boson production was obtained in \refscite{Anastasiou:2015ema,Duhr:2020seh,Mistlberger:2018etf,Anastasiou:2014lda,Anastasiou:2014vaa}.
We confirm that we can reproduce the first term of the threshold expansion of all partonic initial states contributing to the collinear limit of the partonic cross sections using the collinear partonic coefficient functions obtained here after integration over phase space.

\begin{figure*}
 \centering
 \includegraphics[width=0.49\textwidth]{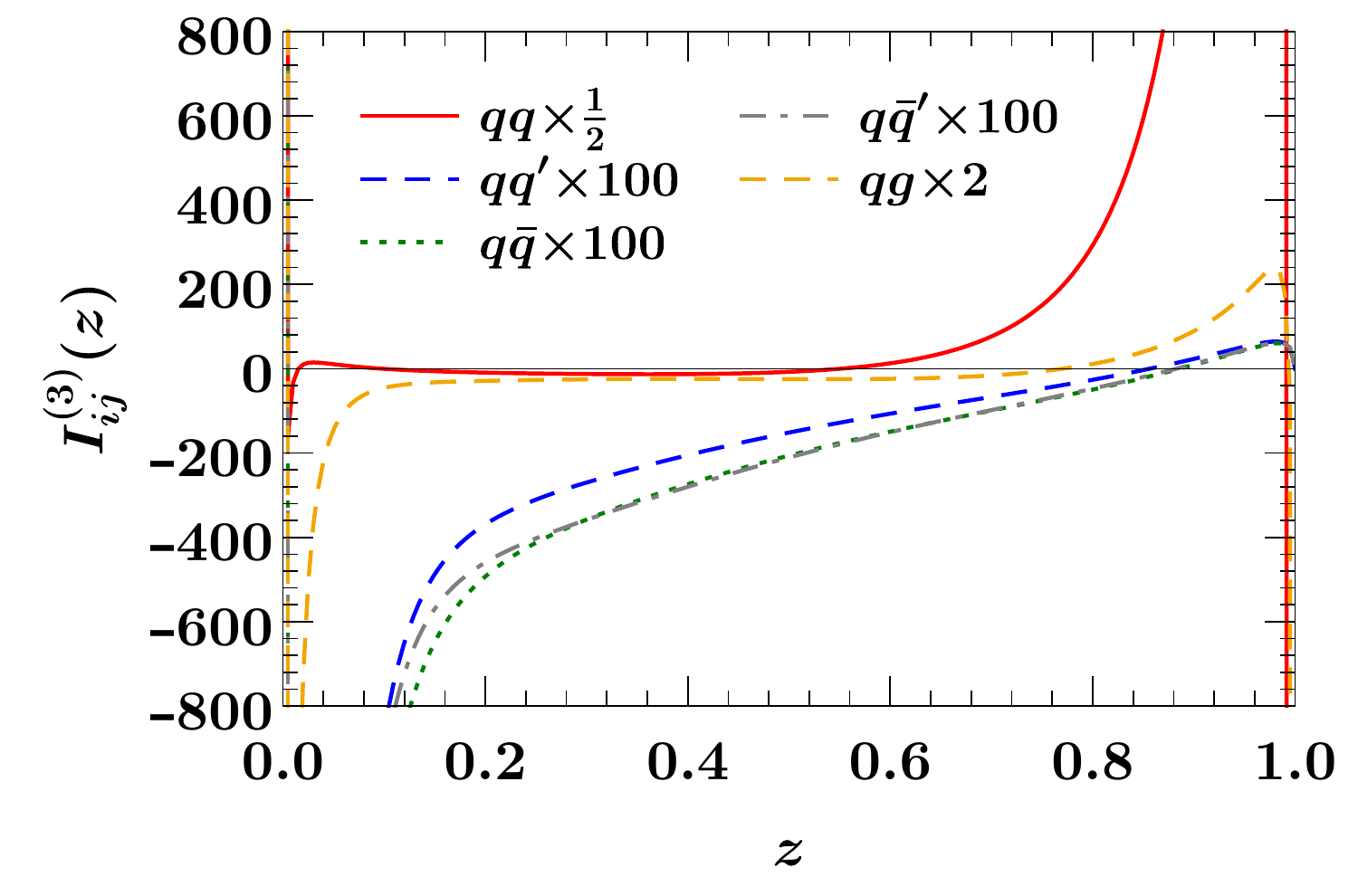}
 \hfill
 \includegraphics[width=0.49\textwidth]{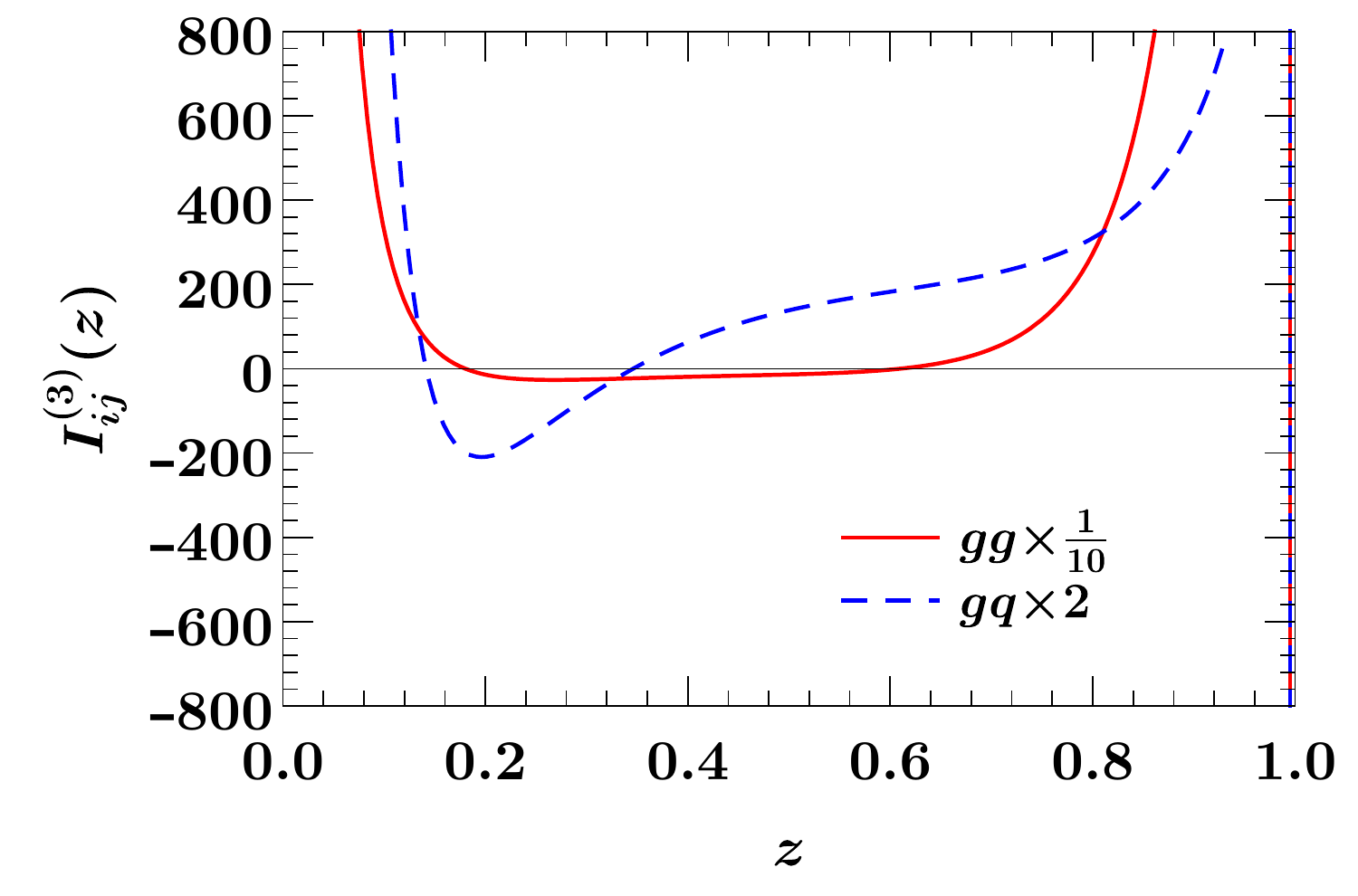}
 \caption{The N$^3$LO boundary term $I^{(3)}_{ij}$ as a function of $z$ in all channels contributing to the quark beam function (left) and the gluon beam function (right). The different channels are rescaled as indicated in the figures.}
 \label{fig:I3}
\end{figure*}

To illustrate our results, \fig{I3} shows the beam function boundary terms $I_{ij}(z)$ relevant for the quark beam function (left) and gluon beam function (right) as a function of $z$. For the purpose of this plot, we replace the occurring distributions as $\delta(1-z) \to 0$ and $\cL_n(1-z) \to \ln^n(1-z) / (1-z)$. Since the different channels give rise to very different shapes and magnitudes, they are rescaled as indicated for illustration purposes only.

To study the impact of our calculation on the beam function itself, we consider the cumulative beam function
\begin{align} \label{eq:cumulative}
 B_i(t_\cut,z,\mu)
 = \int_0^{t_\cut} \df t\, B_i(t,z,\mu)
 = \sum_j \int_z^1 \frac{\df z'}{z'} \int_0^{t_\cut} \df t \, \cI_{ij}(t,z',\mu) f_j\Bigl(\frac{z}{z'},\mu\Bigr)
\,,\end{align}
where we distinguish both quantities only by their arguments. As indicated, this always involves the sum over all flavors $j$ contribution the desired beam function of flavor $i$. We use the \texttt{MMHT2014nnlo68cl} PDF set from \refcite{Harland-Lang:2014zoa} with $\as(m_Z)=0.118$, and evaluate \eq{cumulative} through an implementation of our results in \texttt{SCETlib}~\cite{scetlib}.

\begin{figure*}
 \centering
 \includegraphics[width=0.49\textwidth]{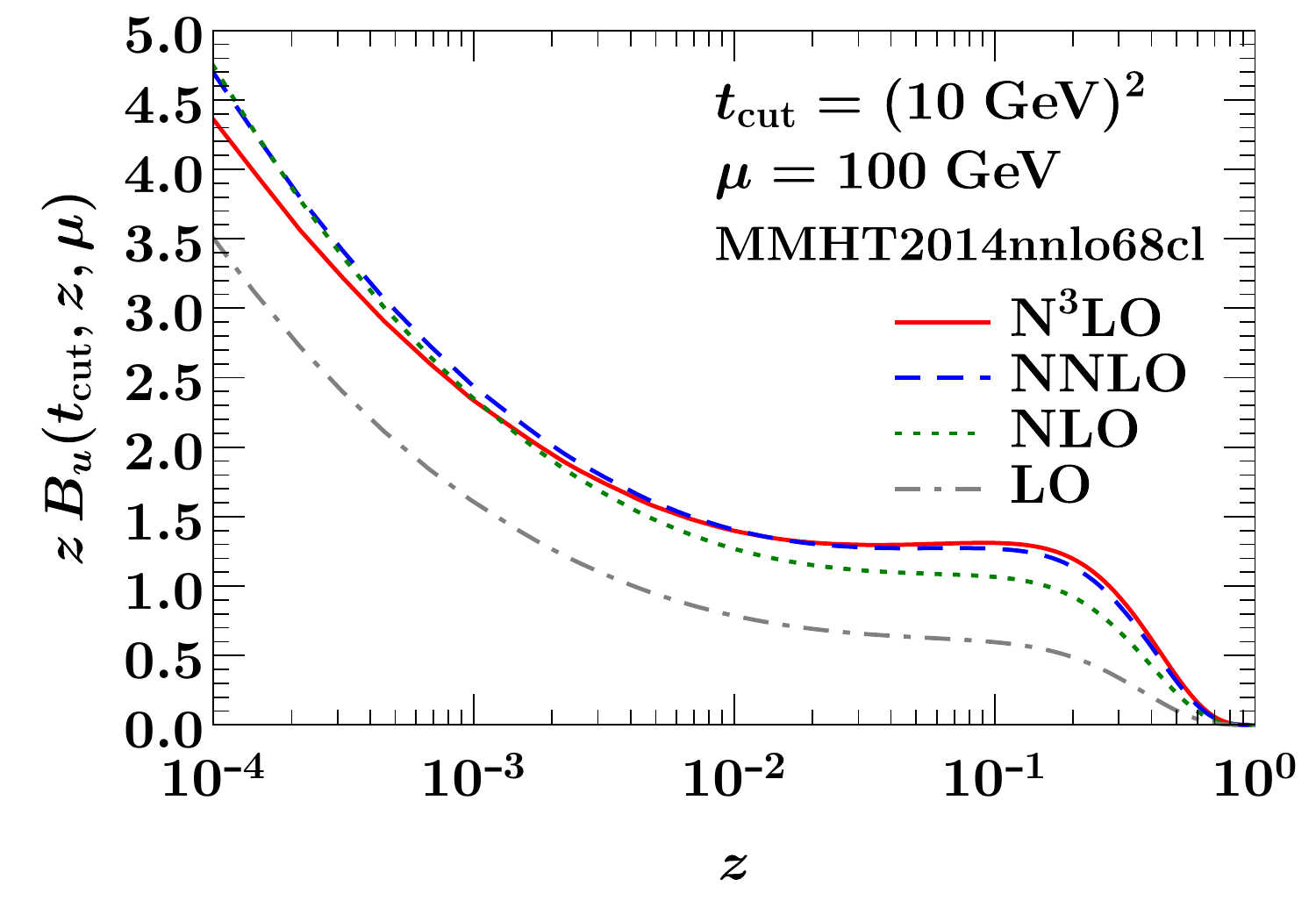}
 \hfill
 \includegraphics[width=0.49\textwidth]{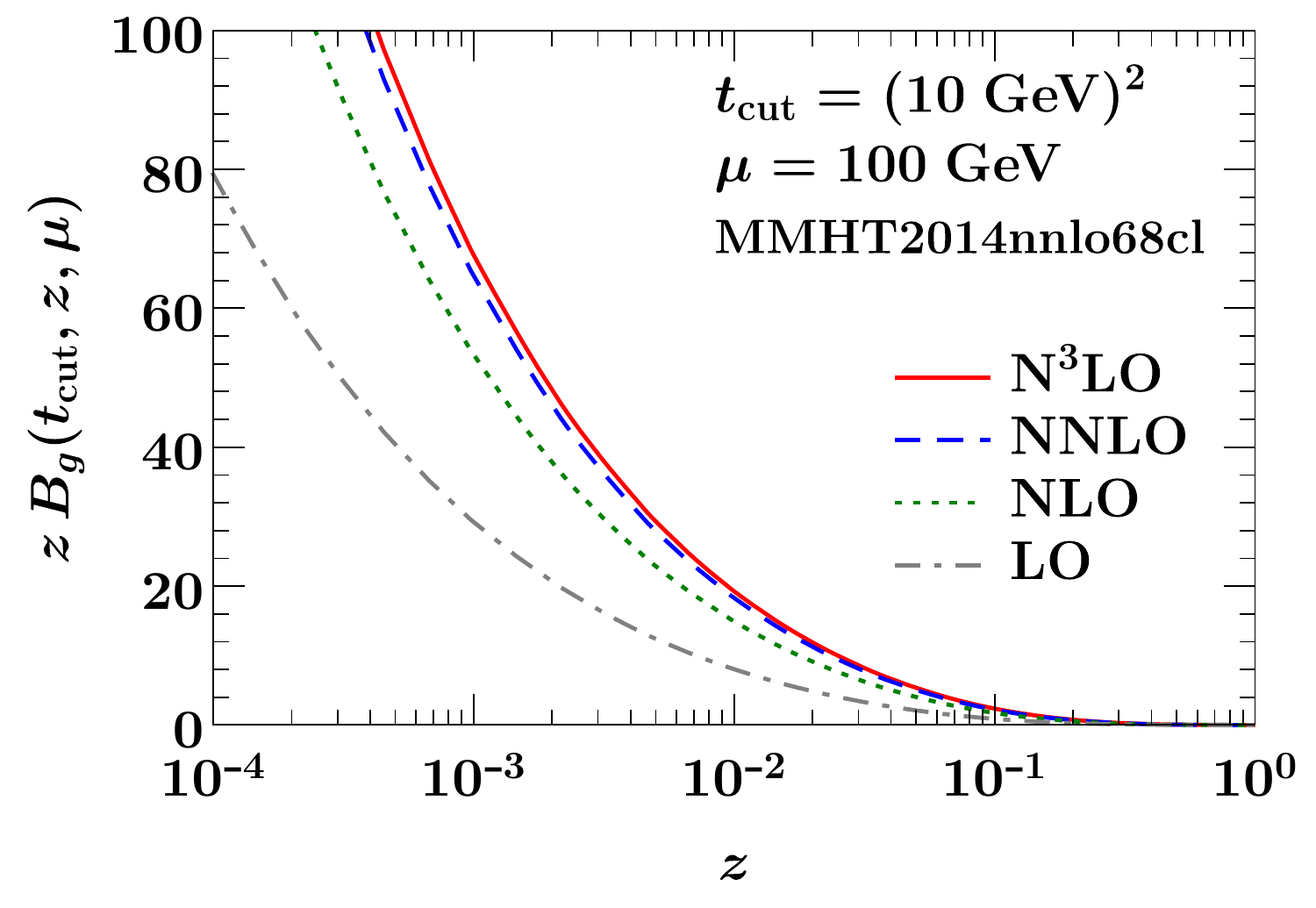}
 \caption{The cumulative $u$-quark (left) and the gluon (right) beam functions as a function of $z$ for fixed $t_{\rm cut} = (10~\GeV)^2$ and $\mu=100~\GeV$. We show the result at LO (which corresponds to the PDF), NLO, NNLO and N$^3$LO.}
 \label{fig:beam_convergence}
\end{figure*}

In \fig{beam_convergence}, we compare the $u$-quark beam function (left) and gluon beam function (right) at LO (gray, dot-dashed), NLO (green, dotted), NNLO (blue, dashed) and N$^3$LO (red, solid) as a function of $z$.
We fix $t_\cut = (10~\GeV)^2$ and $\mu = 100~\GeV$ and rescale the beam functions by $z$.
Note that the LO result corresponds to the PDF itself, and thus illustrates the different shape of the beam function compared to the PDF. While we observe a notable effect of the N$^3$LO corrections, the beam functions show good convergence overall.

\begin{figure*}
 \centering
 \includegraphics[width=0.49\textwidth]{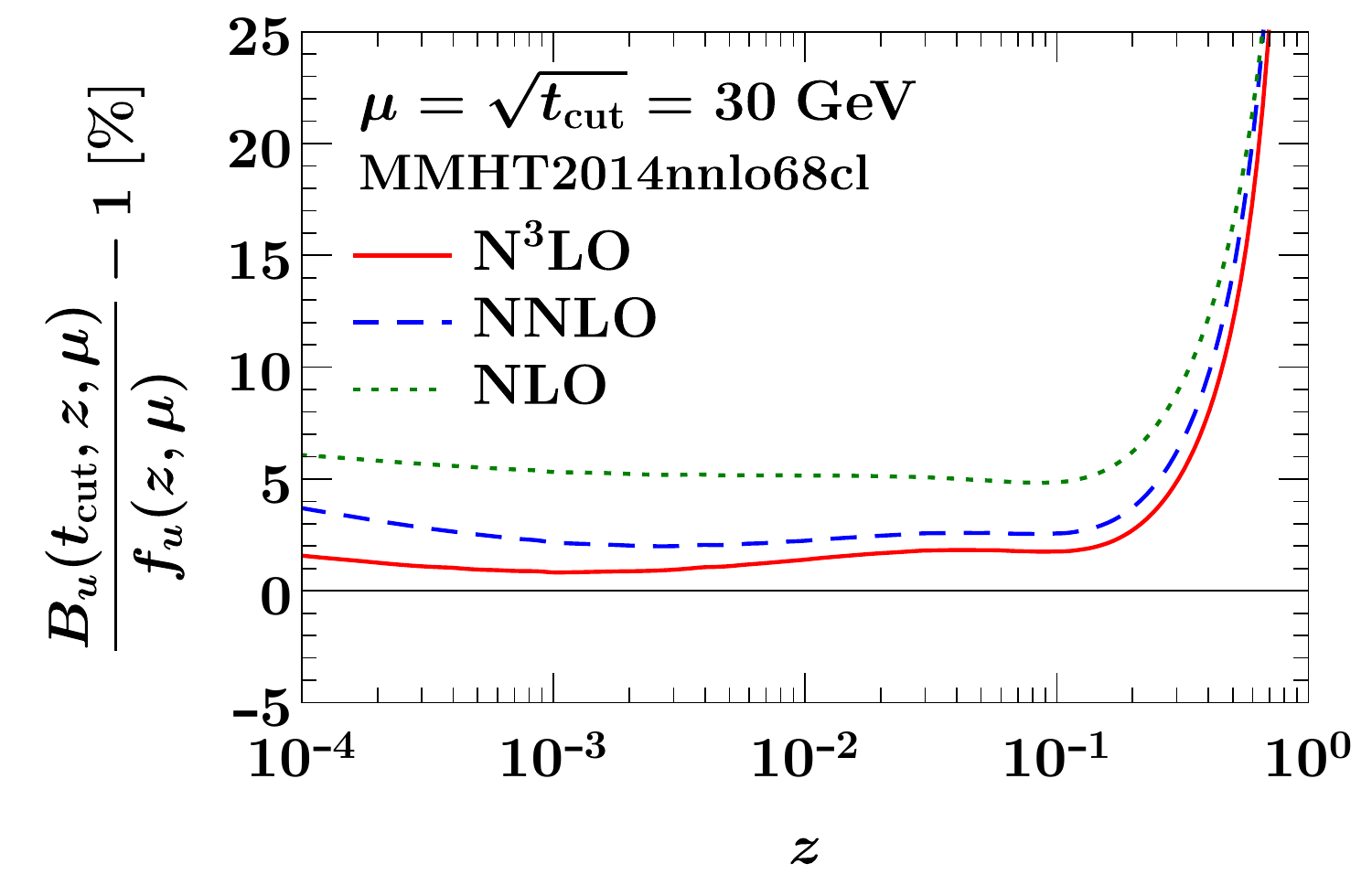}
 \hfill
 \includegraphics[width=0.49\textwidth]{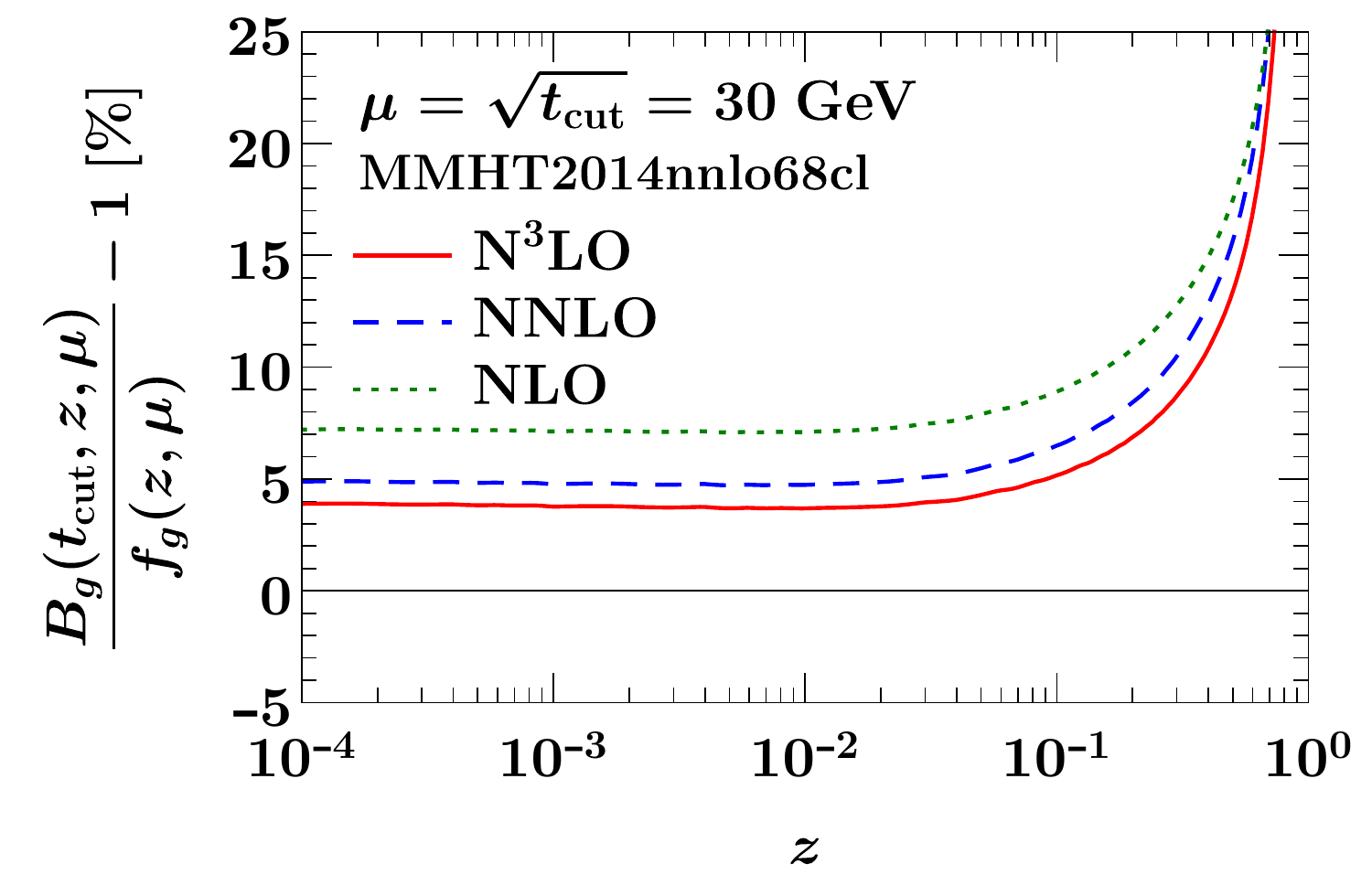}
 \caption{The relative difference of cumulative $u$-quark beam function (left) and the cumulative gluon beam function (right) to the corresponding PDF, as a function. We fix $\mu = \sqrt{t_{\rm cut}} = 30~\GeV$, such that the shown beam function corresponds to the boundary term in a resummed prediction. The different colors show the results at NLO, NNLO and N$^3$LO, respectively.}
 \label{fig:ratio_to_LO}
\end{figure*}

To judge the impact of the new three-loop boundary term $I_{ij}^{(3)}$ on resummed predictions, it is more useful to show the beam function $B_i(t_\cut,z,\mu)$ at its canonical scale $\mu = \sqrt{t_\cut}$, where all distributions $\cL_m$ in \eq{I_log_expansion} vanish and only the boundary term $I_{ij}^{(3)}$ contributes.
In \fig{ratio_to_LO}, we show the cumulative beam functions at the canonical scale with $\mu = \sqrt{t_{\rm cut}} = 30~\GeV$, showing the relative difference of the $u$-quark beam function (left) and the gluon beam function (right) at NLO (green, dotted), NNLO (blue, dashed) and N$^3$LO (red, solid) to the corresponding PDF itself.
We observe that the shape of the beam functions differ significantly from the shape of the PDF for large $z$,
but tend to converge towards the PDF for small $z \lesssim 10^{-1}$.
As before, we see good convergence at N$^3$LO, but still a notable effect of the N$^3$LO corrections itself.

\begin{figure*}
 \centering
 \includegraphics[width=0.6\textwidth]{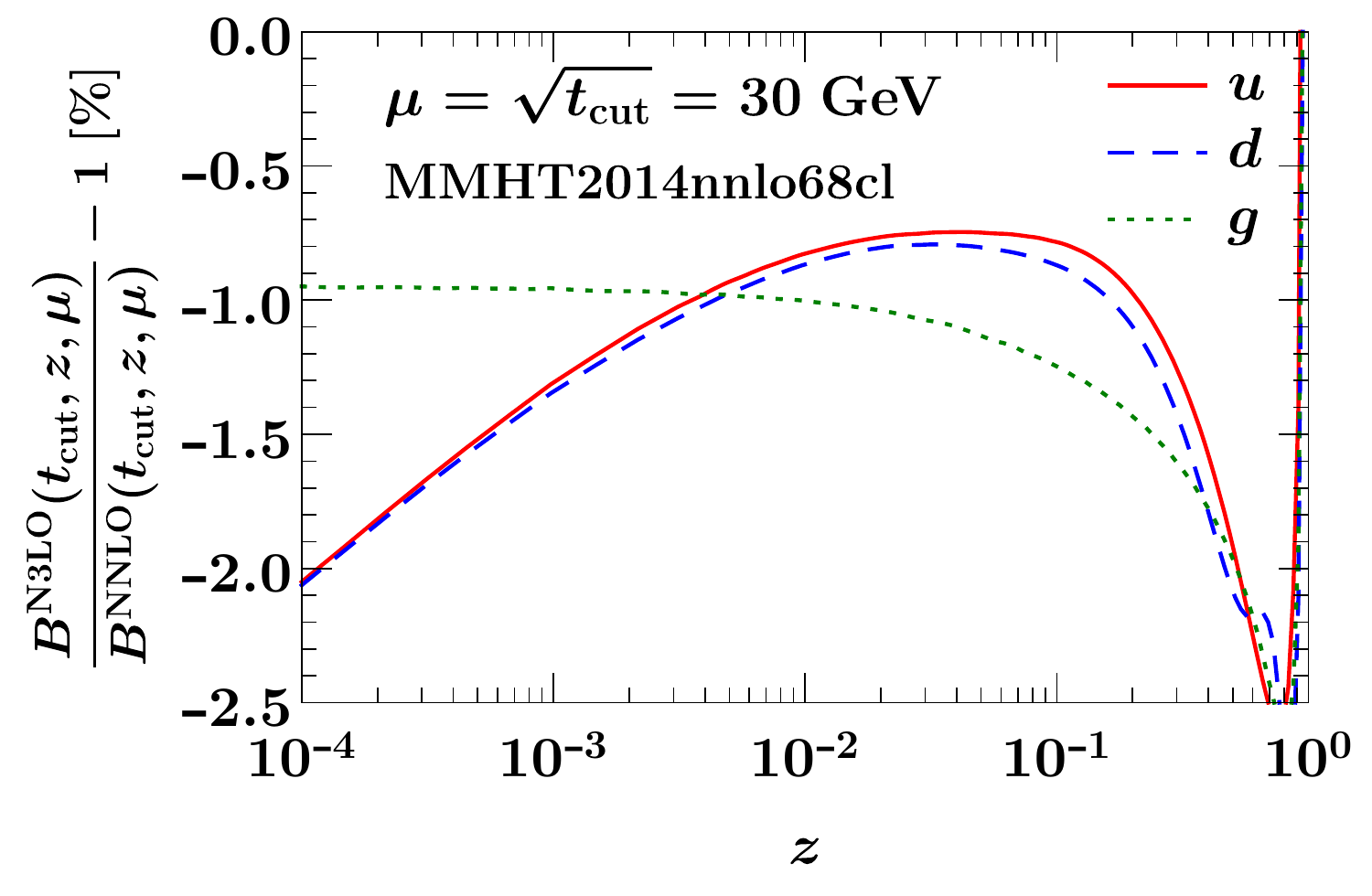}
 \caption{The $K$-factor of the N$^3$LO beam function, i.e.\ the ratio of the N$^3$LO contribution to the NNLO beam function.We fix $\mu = \sqrt{t_{\rm cut}} = 30~\GeV$, such that the shown beam function corresponds to the boundary term in a resummed prediction. The different colors show the results for an $u$-quark, $d$-quark and gluon, respectively.}
 \label{fig:Kfactor}
\end{figure*}

Finally, in \fig{Kfactor} we show the $K$-factor of the N$^3$LO beam function, which we define as the ratio of the N$^3$LO beam to the NNLO beam function.
As before, we choose the canonical scales $\mu = \sqrt{t_{\rm cut}} = 30~\GeV$ as relevant for a resummed calculation, We show the $K$ factor for $u$ quarks (red, solid), $d$ quarks (blue, dashed) and gluons (green, dotted). In all cases, we see corrections of $\sim1-2\%$ with a sizable dependence on $z$.

For completeness, we also show the high-energy limit $z\to0$ of the kernels $I_{ij}^{(3)}(z)$ in \app{high_energy}.
This limit is for example interesting because the small-$\Tau_1$ region is known to grow at small $z$ in deep inelastic scattering~\cite{Kang:2013nha,Kang:2014qba}.

\section{Conclusions}
\label{sec:conclusion}

We have calculated the perturbative matching kernel relating $N$-jettiness beam functions with lightcone PDFs in all partonic channels for the first time at N$^3$LO in QCD.
Our calculation is based on a method recently developed by us to expand hadronic collinear cross sections~\cite{Ebert:CollExp}, demonstrating its usefulness for the calculation of universal ingredients arising in the collinear limit of QCD.

We provide our results in the form of ancillary file with this submission, where we include the renormalized $N$-jettiness beam function in both momentum ($t$) and Fourier ($y$) space.
For the $t$ space result, we also provide its expansions around $z=0$ and $z=1$ through 50 orders in the expansion.

In contrast to the TMD beam functions, which are based on the same collinear limit and at N$^3$LO can be entirely expressed in terms of harmonic polylogarithms up to weight 5~\cite{Luo:2019szz,Ebert:PTBF}, the $\Tau_N$ beam functions have a much richer structure of the appearing functions and are expressed in terms of Goncharov polylogarithm, as well as iterated integrals with letters that involve square roots.
It will be interesting to better understand the source of this difference.

Our results have various phenomenological applications.
Firstly, we provide a key ingredient to extend the $N$-jettiness subtraction method~\cite{Boughezal:2015dva,Gaunt:2015pea} to N$^3$LO, which can be used to obtain exact fully-differential cross sections at this order.
They are also crucial to extend the resummation of $\Tau_N$ to N$^3$LL$^\prime$ and N$^4$LL accuracy, and for matching N$^3$LO calculations to parton showers based on $\Tau_0$ resummation~\cite{Alioli:2012fc,Alioli:2015toa}.

It will also be interesting to further study the collinear limit of QCD using the underlying method of collinear expansions.
In particular, we expect this to shed light on the universal structure of $\Tau_N$ factorization at subleading power \cite{Kolodrubetz:2016uim,Feige:2017zci,Moult:2017rpl,Chang:2017atu,Beneke:2017ztn,Beneke:2018rbh,Moult:2018jjd,Moult:2019mog,Moult:2019uhz}, which has recently attracted much attention in the literature due to its importance for $\Tau_N$-subtractions~\cite{Moult:2016fqy,Boughezal:2016zws,Moult:2017jsg,Boughezal:2018mvf,Ebert:2018lzn,Bhattacharya:2018vph,Boughezal:2019ggi,Ebert:2019zkb}.

\acknowledgments
We thank Johannes Michel, Iain Stewart and Frank Tackmann for useful discussions.
This work was supported by the Office of Nuclear Physics of the U.S.\ Department of Energy under Contract No.\ DE-SC0011090 and DE-AC02-76SF00515.
M.E.\ is also supported by the Alexander von Humboldt Foundation through a Feodor Lynen Research Fellowship,
and B.M.\ is also supported by a Pappalardo fellowship.

\appendix

\section{Ingredients for the calculation of the beam function}
\label{app:details}

In this appendix, we provide more details on the regularization and renormalization of the beam function kernels. Details of the calculation of all required integrals will be presented in \refcite{Ebert:ThingsToCome}.

\subsection{Renormalization group equations}
\label{app:RGE}

In $t$ space, the beam function $B_i(t,z,\mu)$ obeys the RGE~\cite{Stewart:2009yx, Stewart:2010qs}
\begin{align} \label{eq:beam_RGE}
 \mu \frac{\df}{\df \mu} B_i(t, z, \mu) &= \int\! \df t'\, \gamma_B^i(t-t',\mu)\, B_i(t', z, \mu)
\,,\end{align}
where the anomalous dimension $\gamma_B^i$ has the all-order form
\begin{align} \label{eq:gamma_B}
 \gamma_B^i(t, \mu) = -2 \GammaC^i[\as(\mu)]\,\cL_0(t, \mu^2) + \gamma_B^i[\as(\mu)] \, \delta(t)
\,.\end{align}
Here, $\GammaC^i(\as)$ and $\gamma^i_B(\as)$ are the cusp and beam noncusp anomalous dimensions,
which both depend on the color representation $i=q$ or $i=g$ only, but are independent of the quark flavor.
The RGE for the matching kernel follows from \eqs{beam_OPE}{beam_RGE} and the DGLAP equation
\begin{equation} \label{eq:DGLAP}
 \mu \frac{\df}{\df\mu} f_i(z,\mu)
 = 2 \sum_j \int_z^1\!\frac{\df z'}{z'}\, P_{ij}(z',\mu)\, f_j\Bigl(\frac{z}{z'}, \mu\Bigr)
\,.\end{equation}
It is given by \cite{Stewart:2010qs}
\begin{equation} \label{eq:I_RGE}
 \mu\frac{\df}{\df\mu}\cI_{ij}(t,z,\mu)
= \sum_k \int\!\df t'\, \int_z^1\!\frac{\df z'}{z'}\, \cI_{ik}\Bigl(t - t', \frac{z}{z'},\mu\Bigr)
  \Bigl[ \gamma_B^i(t', \mu)\, \delta_{kj} \delta(1-z') - \delta(t')\, 2P_{kj}(z',\mu) \Bigr]
\,.\end{equation}

\subsection{Structure of the beam function counterterm}
\label{app:counter_term}

We define the Fourier transformation of a function $f$ as
\begin{align}
 \tilde f(y,\cdots) = \int\df t \, e^{-\img t y} \, f(t,\cdots)
\,,\qquad
 f(t,\cdots) = \int\frac{\df y}{2\pi} \, e^{\img t y} \, \tilde f(y,\cdots)
\,.\end{align}
The Fourier transform of the bare kernel $\cI_{ij}(t,z,\eps)$ can be conveniently evaluated using
\begin{align} \label{eq:FT_tau}
 \mu^{2a\eps} \int_{-\infty}^\infty \df t \,e^{-\img t y} \frac{\theta(t)}{t^{1 + a \eps}} = e^{a \eps (L_y - \gamma_E)} \Gamma(-a\eps)
\,,\qquad
 L_y = \ln\bigl(\img y \mu^2 e^{\gamma_E}\bigr)
\,.\end{align}
Here, $L_y$ is the canonical logarithm in Fourier space, and $\gamma_E$ is the Euler-Mascheroni constant.
In Fourier space, the renormalization of the bare matching kernel in \eq{Iij_ren} becomes multiplicative in $y$,
\begin{equation} \label{eq:Iij_ren_FT}
 \tilde \cI_{ij}(y,z,\mu)
 = \sum_{k} \int_z^1\!\frac{\df z'}{z'} \, \Gamma_{jk}\Bigl(\frac{z}{z'},\eps\Bigr) \,
   \tilde Z_B^i(y,\eps,\mu)  \hat Z_{\as}(\mu,\eps) \, \tilde\cI_{i k}(y,z',\eps)
\,,\end{equation}
and the counterterm $\tilde Z_B^i$ follows from the RG \eq{gamma_B} in $y$ space,
\begin{align} \label{eq:TBF_RGE_FT}
 \frac{\df}{\df\ln\mu} \ln{\tilde B_i(y,z,\mu)} &
 = \tilde\gamma_B^i(y,\mu)
 = - \frac{\df}{\df\ln\mu} \ln \tilde Z_B^i(y,\mu,\eps)
 \nn\\&
 = 2 \GammaC^i[\as(\mu)] L_y + \gamma_B^i[\as(\mu)]
\,.\end{align}
Solving \eq{TBF_RGE_FT}, we can predict the all-order pole structure of $\tilde Z_B^i$ as (see also \refcite{Becher:2009cu})
\begin{align} \label{eq:ZB_TBF}
 \ln \tilde Z^B_i(y, \mu, \eps) = -\int\limits_0^{\as(\mu)} \!\! \frac{\df\alpha}{\beta(\alpha,\eps)} \biggl[
  4 \GammaC^i(\alpha) \int_{\as(\mu)}^{\alpha} \frac{\df\alpha'}{\beta(\alpha',\eps)}
  + 2 \GammaC^i(\alpha) L_y
  + \gamma_B^i(\alpha) \biggr]
\,,\end{align}
where $\beta(\alpha_s,\eps) = -2 \eps \alpha_s + \beta(\alpha_s)$ is the QCD beta function in $d=4-2\eps$ dimensions. Expanding \eq{ZB_TBF} systematically in $\alpha$, we obtain the result through three loops as
\begin{align} \label{eq:TBF_Z_2}
 \ln \tilde Z^B_i(y, \mu, \eps) &=
 \frac{\as}{4\pi} \biggl\{
 \frac{\Gamma_0^i}{\eps^2}+\frac{1}{2\eps} \bigl(2\Gamma_0^i L_y  + \gamma_{B\,0}^i\bigr)
 \biggr\}
\nn\\&
 + \Bigl(\frac{\as}{4\pi}\Bigr)^2 \biggl\{
   -\frac34 \frac{\beta_0 \Gamma_0^i}{\eps^3}
   - \frac{1}{4\eps^2} \bigl[ \beta_0 \bigl(2 \Gamma_0^i L_y + \gamma_{B\,0}^i \bigr) - \Gamma_1^i \bigr]
   + \frac{1}{4\eps} \bigl(2 \Gamma_1^i L_y + \gamma_{B\,1}^i\bigr)
 \biggr\}
\nn\\&
 + \Bigl(\frac{\as}{4\pi}\Bigr)^3 \biggl\{
   \frac{11}{18} \frac{\beta_0^2 \Gamma_0^i}{\eps^4}
   +\frac{1}{6 \eps^3}\biggl[ \beta_0^2  \bigl( 2 \Gamma_0^i L_y + \gamma_{B\,0}^i\bigr) - \frac{5}{3} \beta_0 \Gamma_1^i - \frac{8}{3} \beta_1 \Gamma_0^i \biggr]
   \nn\\&\hspace{1.8cm}
   - \frac{1}{6\eps^2} \biggl[ \beta_1 \bigl( 2 \Gamma_0^i L_y  + \gamma_{B\,0}^i\bigr) + \beta_0 \bigl( 2 \Gamma_1^i L_y  + \gamma_{B\,1}^i \bigr) - \frac23 \Gamma_2^i \biggr]
   \nn\\&\hspace{1.8cm}
   +\frac{1}{6\eps} \bigl(2 \Gamma_2^i L_y  + \gamma_{B\,2}^i \bigr)
 \biggr\}
 + \cO(\as^4)
\,.\end{align}
Here, the $\gamma_n$ are the coefficients of the corresponding anomalous dimensions at $\cO[(\as/4\pi)^n]$. 
Explicit expressions for all anomalous dimensions in the convention of \eq{TBF_Z_2} are collected in~\refcite{Billis:2019vxg}.
The required three-loop results for $\GammaC$ and $\beta$ were calculated in \refscite{Korchemsky:1987wg, Moch:2004pa, Vogt:2004mw} and \refscite{Tarasov:1980au, Larin:1993tp}, respectively.
The beam noncusp anomalous dimension were originally determined in \refscite{Stewart:2010qs,Berger:2010xi}, see also \refscite{Bruser:2018rad,Banerjee:2018ozf}.

\subsection[\texorpdfstring{$\as$}{alphaS} renormalization and IR counterterms]
           {\boldmath $\as$ renormalization and IR counterterms}
\label{app:UV_IR_counter_terms}

The bare strong coupling constant is renormalised as
\bea
\as^b&=&\as\left(\frac{\mu^2}{4\pi} e^{\gamma_E}\right)^{\eps} \left[1+\frac{\as}{4\pi}\left(-\frac{\beta_0}{\epsilon }\right)+\left(\frac{\as}{4\pi}\right)^2\left(\frac{\beta_0^2}{\epsilon ^2}-\frac{\beta_1}{2 \epsilon }\right)\nn\right.\\
&&\hspace{2.5cm}+\left.\left(\frac{\as}{4\pi}\right)^3\left(-\frac{\beta_0^3}{\epsilon ^3}+\frac{7 \beta_1 \beta_0}{6 \epsilon ^2}-\frac{\beta_2}{3 \epsilon }\right)+\mathcal{O}(\as^4)\right]\,.
\eea
The mass factorisation counter term can be expressed in terms of the splitting functions $P_{ij}$~\cite{Moch:2004pa,Vogt:2004mw} as
\bea \label{eq:Gamma}
\Gamma_{ij}(z) =&& \delta_{ij}\delta(1-z) \nn\\
&&+\left(\frac{\as}{4\pi}\right)\frac{P^{(0)}_{ij}}{\epsilon} \nn\\
&&+\left(\frac{\as}{4\pi}\right)^2\left[\frac{1}{2\epsilon^2}\left(P^{(0)}_{ik}\otimes P^{(0)}_{kj}-\beta_0  P^{(0)}_{ij}\right)+\frac{1}{2\epsilon}P^{(1)}_{kj}\right]\nn\\
&&+\left(\frac{\as}{4\pi}\right)^3\left[\frac{1}{6\epsilon^3}\left(P^{(0)}_{ik}\otimes P^{(0)}_{kl}\otimes P^{(0)}_{lj}-3\beta_0P^{(0)}_{ik}\otimes P^{(0)}_{kj}+2\beta_0^2P^{(0)}_{ij}\right)\right.\nn\\
&&\hspace{1.6cm}\left.+\frac{1}{6\epsilon^2}\left(P^{(1)}_{ik}\otimes P^{(0)}_{kj}+2P^{(0)}_{ik}\otimes P^{(1)}_{kj}-2\beta_0 P^{(1)}_{ij}-2 \beta_1 P^{(0)}_{ij}\right)+\frac{1}{3\epsilon}P^{(2)}_{ij}\right].\nn\\
&&+~\mathcal{O}(\as^4)
\,.\eea
Here, we suppress the argument $z$ of the splitting functions on the right hand side and keep the summation over repeated flavor indices implicit. The convolution in \eq{Gamma} is defined as
\beq
f\otimes g=\int_z^1\frac{\df z^\prime}{z^\prime} f(z)g\left(\frac{z}{z^\prime}\right).
\eeq

\section{High-energy limit of the beam function kernels}
\label{app:high_energy}

Here, we present the high-energy limit $z\to0$ of the beam function $I_{ij}^{(3)}(z)$:
{
\allowdisplaybreaks
\bea
\lim\limits_{z\rightarrow0} z \, I_{gg}^{(3)}(z) &=&
-\frac{1}{120} C_A^3 \ln^5(z) + \ln^4(z) \left(\frac{11 C_A^3}{72}-\frac{C_A^2 n_f}{72}-\frac{C_A C_F n_f}{72}\right)
\nn\\&+&
\ln^3(z) \left[C_A^3 \left(-\frac{\zeta_2}{4}-\frac{229}{216}\right)+\frac{17 C_A^2 n_f}{54}+C_A \left(\frac{C_F n_f}{6}-\frac{n_f^2}{108}\right)-\frac{C_F n_f^2}{27}\right]
\nn\\&+&
\ln^2(z) \left[C_A^3 \left(\frac{143 \zeta_2}{16}-\frac{25 \zeta_3}{24}-\frac{1013}{96}\right)+C_A^2 n_f \left(\frac{1207}{432}-\frac{17 \zeta_2}{24}\right)
\right.\nn\\&&\left.\hspace{1cm}+
C_A \left(C_F n_f \left(\frac{745}{864}-\frac{3 \zeta_2}{4}\right)-\frac{23 n_f^2}{216}\right)+C_F^2 n_f \left(\frac{11}{48}-\frac{\zeta_2}{6}\right)-\frac{5 C_F n_f^2}{27}\right]
\nn\\&+&
\ln(z) \left[C_A^3 \left(\frac{15143 \zeta_2}{432}+\frac{407 \zeta_3}{36}-\frac{1433 \zeta_4}{48}-\frac{43393}{2592}\right) +C_F n_f^2 \left(\frac{\zeta_2}{9}-\frac{25}{81}\right)
\right.\nn\\&&\left.\hspace{1cm}+
C_A C_F n_f \left(-\frac{377}{216} \zeta_2-\frac{22 \zeta_3}{9}+\frac{5033}{3888}\right)+C_A n_f^2 \left(\frac{\zeta_2}{18}-\frac{251}{648}\right)
\right.\nn\\&&\left.\hspace{1cm}+
C_F^2 n_f \left(\frac{311}{288}-\frac{25 \zeta_2}{36}\right)+C_A^2 n_f \left(-\frac{1031}{216} \zeta_2-\frac{\zeta_3}{18}+\frac{11027}{1296}\right)\right]
\nn\\&+&
\cO(\ln^0 z)
\,,\\
\lim\limits_{z\rightarrow0} z \, I_{gq}^{(3)}(z) &=&
-\frac{1}{120} C_A^2 C_F \ln^5(z)+\ln^4(z) \left(\frac{5 C_A^2 C_F}{36}-\frac{C_A C_F n_f}{36}+\frac{C_F^2 n_f}{72}\right)
\nn\\&+&
\ln^3(z) \left[C_A^2 C_F \left(\frac{\zeta_2}{12}-\frac{89}{72}\right)+C_A C_F^2 \left(\frac{5}{24}-\frac{\zeta_2}{3}\right)
\right.\nn\\&&\left.\hspace{1cm}
+C_AC_F n_f \frac{97}{216}-\frac{C_F^2 n_f}{54}-\frac{C_F n_f^2}{36}\right]
\nn\\&+&
\ln^2(z) \left[C_A^2 C_F \left(\frac{103 \zeta_2}{16}-\frac{7 \zeta_3}{24}-\frac{83}{9}\right)+C_A C_F^2 \left(\frac{73 \zeta_2}{24}-\frac{5 \zeta_3}{4}-\frac{151}{96}\right)
\right.\nn\\&&\left.\hspace{1cm}
- \frac{5 C_F n_f^2}{36}
+ C_A C_F n_f \left(\frac{2275}{864}-\frac{2 \zeta_2}{3}\right)+C_F^3 \left(-\frac{3 \zeta_2}{4} +\frac{\zeta_3}{2}+\frac{13}{16}\right)
\right.\nn\\&&\left.\hspace{1cm}
+C_F^2 n_f \left(\frac{157}{432}-\frac{3 \zeta_2}{4}\right)\right]
\nn\\&+&
\ln (z) \left[C_F^3 \left(-\frac{19}{8}  \zeta_2+\frac{3 \zeta_3}{2}-5 \zeta_4+\frac{93}{16}\right)+C_A C_F n_f \left(-\frac{265}{108}  \zeta_2-\frac{\zeta_3}{6}+\frac{1619}{324}\right)
\right.\nn\\&&\left.\hspace{1cm}+
C_F^2 n_f \left(-\frac{193}{108} \zeta_2-\frac{37 \zeta_3}{18}+\frac{24757}{7776}\right)
\right.\nn\\&&\left.\hspace{1cm}
+ C_A C_F^2 \left(\frac{757 \zeta_2}{48}-\frac{9 \zeta_3}{4}-\frac{45 \zeta_4}{8}-\frac{3055}{192}\right)
\right.\nn\\&&\left.\hspace{1cm}+
C_A^2 C_F \left(\frac{2099 \zeta_2}{108}+\frac{106 \zeta_3}{9}-\frac{923 \zeta_4}{48}-\frac{3377}{576}\right) -\frac{25 C_F n_f^2}{108}
\right]
\nn\\&+&
 \cO(\ln^0 z)
\,,\\
\lim\limits_{z\rightarrow0} z \, I_{qg}^{(3)}(z) &=&
C_A^2 \left(\frac{2 \zeta_3}{9}-\frac{322}{243}\right) \ln (z)
\nn\\&+&
C_A^2 \left(-\frac{1}{324} 337 \zeta_2+\frac{787 \zeta_3}{432}+\frac{263 \zeta_4}{144}-\frac{266675}{23328}\right)+ C_A n_f \left(-\frac{\zeta_3}{27}-\frac{1169}{23328}\right)
\nn\\&+&
C_AC_F \left(-\frac{\zeta_2}{108}-\frac{7 \zeta_4}{6}-\frac{\zeta_3}{12}-\frac{1103}{1728}\right)+C_F n_f \left(\frac{6049}{11664}-\frac{2 \zeta_3}{27}\right)
\,,\\
\lim\limits_{z\rightarrow0} z \, I_{qq}^{(3)}(z)
&=& \lim\limits_{z\rightarrow0} z \, I_{q\bar q}^{(3)}(z)
 = \lim\limits_{z\rightarrow0} z \, I_{q q^\prime}^{(3)}(z)
 =\lim\limits_{z\rightarrow0} z \,  I_{q \bar{q}^\prime}^{(3)}(z) \nn\\
&=&
C_A C_F \left(\frac{2 \zeta_3}{9}-\frac{322}{243}\right) \ln (z)
+C_F^2 \left(-\frac{\zeta_2}{108}-\frac{7 \zeta_4}{6}-\frac{\zeta_3}{12}-\frac{1103}{1728}\right)
\nn\\
&+& C_F n_f \left(\frac{305}{1458}-\frac{2 \zeta_3}{27}\right)
+ C_A C_F \left(-\frac{337}{324} \zeta_2+\frac{257 \zeta_3}{144}+\frac{263 \zeta_4}{144}-\frac{258211}{23328}\right)
\!.\eea
Here, the color factors $C_A$ and $C_F$ are only used for compactness of the result and should be replaced with their expressions in terms of $n_c$.
Note that the expressions for the high energy limit ${z\rightarrow 0}$ up to $\cO(z^{50})$, as well as that for the threshold limit $z\to 1$ up to $\cO((1-z)^{50})$, can be found in electronic form in the ancillary files.
}

\addcontentsline{toc}{section}{References}
\bibliographystyle{jhep}
\bibliography{TBF.bbl}

\end{document}